\documentclass[conference]{IEEEtran}
\IEEEoverridecommandlockouts
\usepackage{cite}
\usepackage{amsmath,amssymb,amsfonts}
\usepackage{graphicx}
\usepackage{textcomp}
\usepackage{xcolor}
\def\BibTeX{{\rm B\kern-.05em{\sc i\kern-.025em b}\kern-.08em
    T\kern-.1667em\lower.7ex\hbox{E}\kern-.125emX}}

\usepackage[colorlinks=true]{hyperref}
\usepackage{algorithm}
\usepackage{algorithmic}
\usepackage{multirow}
\usepackage{subfig}

\begin{document}

\newcommand{\reffig}[1]{Figure~\ref{#1}}
\newcommand{\reftab}[1]{Table~\ref{#1}}
\newcommand{\refsec}[1]{Section~\ref{#1}}
\newcommand{\refequ}[1]{Equation~\ref{#1}}
\newcommand{\etal}{et~al.~}
\newcommand{\ang}{$\text{\AA}$}
\newcommand{\schr}{Schr\"{o}dinger}
\renewcommand{\algorithmicrequire}{\textbf{Input:}}

\title{Automatic Algorithm Switching for Accurate Quantum Chemical Calculations}

\author{\IEEEauthorblockN{Satoshi Imamura}
\IEEEauthorblockA{s-imamura@fujitsu.com}
\and
\IEEEauthorblockN{Akihiko Kasagi}
\IEEEauthorblockA{kasagi.akihiko@fujitsu.com\\
\textit{Computing Laboratry, Fujitsu Limited, Kawasaki, Japan}}
\and
\IEEEauthorblockN{Eiji Yoshida}
\IEEEauthorblockA{yoshida.eiji-01@fujitsu.com}
}

\maketitle
\thispagestyle{plain}
\pagestyle{plain}

\begin{abstract}
Quantum chemical calculations (QCC) are computational techniques to analyze the characteristics of molecules. The variational quantum eigensolver (VQE) designed for noisy intermediate-scale quantum (NISQ) computers can be used to calculate the ground-state energies of molecules, while coupled-cluster with singles, doubles, and perturbative triples [CCSD(T)] is regarded as a traditional gold standard algorithm in QCC. The advantage between CCSD(T) and VQE in terms of the accuracy of ground-state energy calculation differs depending on molecular structures.

In this work, we propose an automatic algorithm switching (AAS) technique to accurately calculate the ground-state energies of a target molecule with different bond distances. It automatically switches CCSD(T) and VQE by identifying a bond distance where the accuracy of CCSD(T) begins to drop for a target molecule. Our evaluation using a noise-less quantum computer simulator demonstrates that AAS improves the accuracy to describe the bond breaking processes of molecules compared to CCSD(T) and VQE.

\end{abstract}


\section{Introduction} \label{sec:introduction}

Quantum chemical calculations (QCC) are computational techniques to analyze the characteristics and structures of molecules, which have been applied in drug discovery and material development~\cite{Matta:2007qu, Bochevarov:2013ja}. It is a basis task in QCC to calculate the ground-state energies of molecules based on the electronic \emph{\schr} equation, because the characteristic of a target molecule can be analyzed based on the ground-state energies with its different structures.

Coupled-cluster with singles, doubles, and perturbative triples [CCSD(T)] is well known as a traditional gold standard algorithm in QCC because of its high accuracy with moderate computational cost~\cite{Ku:2019ac}. It calculates an electron correlation energy by explicitly treating single and double excitations through an iterative process~\cite{Solomonik:2014ma} and adding the approximate treatment of triple excitations based on the perturbation theory.

Recently, noisy intermediate-scale quantum (NISQ) computers have been developed intensively~\cite{IBM_QCRoadmap:2022, rigetti_QCRoadmap:2022}, and variational quantum eigensolver (VQE) is regarded as one of the most promising algorithms operating on NISQ computers. VQE is a variational hybrid quantum-classical eigensolver and can be used to calculate the ground-state energies of molecules~\cite{Peruzzo:2014va, Tilly:2022th}. It iteratively executes a parameterized quantum circuit on a quantum device and updates its parameters using a classical optimizer until the energy calculated from the measurement result of the quantum circuit converges. The accuracy of VQE strongly depends on an \emph{ansatz}; especially, the unitary coupled-cluster singles and doubles (UCCSD) is a well-known chemistry-inspired ansatz which can achieve a high accuracy in QCC~\cite{Ku:2019ac,Gonthier:2022me}.

The advantage between CCSD(T) and VQE in terms of the accuracy of ground-state energy calculation differs depending on molecular structures. CCSD(T) is basically accurate for stable structures where electron correlation is weak, but not for unstable structures where electron correlation is strong~\cite{Kowalski:2000re}. In contrast, VQE with the UCCSD ansatz (UCCSD-VQE) can represent strong electron correlation accurately~\cite{Culpitt:2023un}.

\begin{figure}[t]
    \centering
    \includegraphics[width=0.8\linewidth]{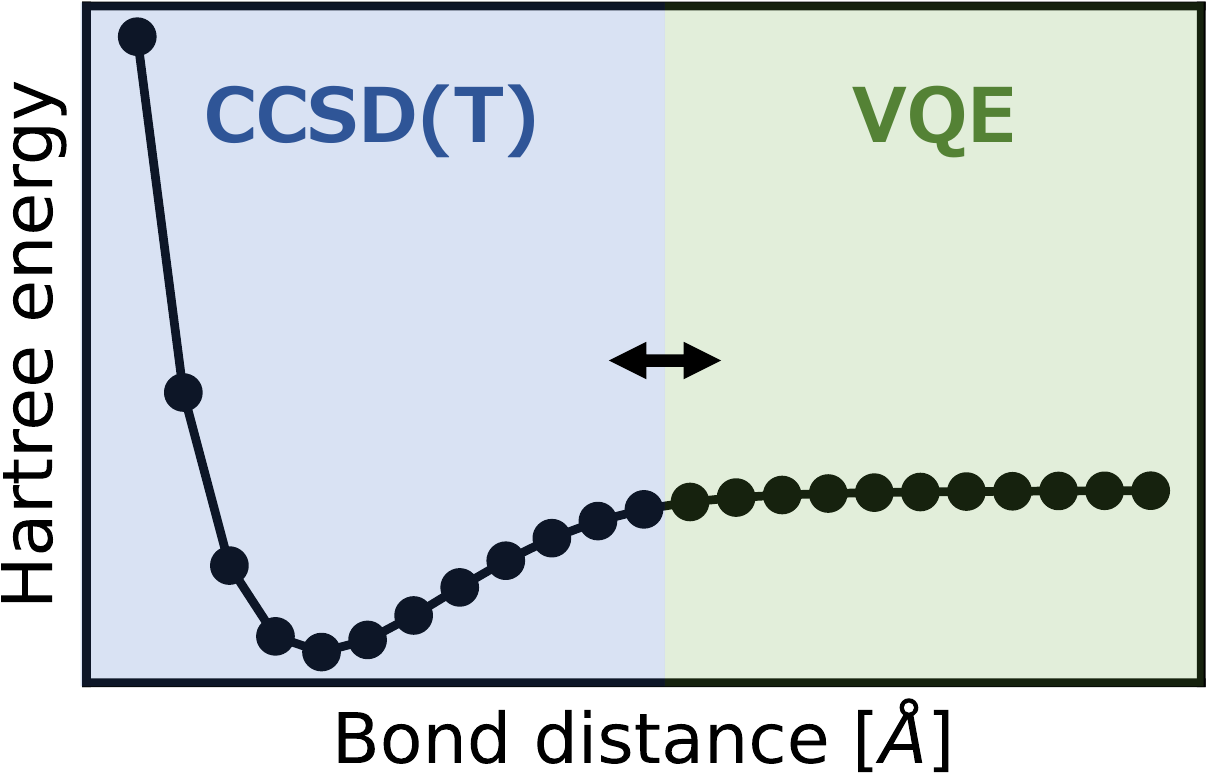}
    \caption{The overview of our proposed automatic algorithm switching (AAS) technique.}
    \label{fig:aas_overview}
\end{figure}

In this work, we propose an \emph{automatic algorithm switching (AAS)} technique to accurately calculate the ground-state energies of a target molecule with different bond distances. As shown in \reffig{fig:aas_overview}, it automatically switches CCSD(T) and VQE according to bond distances by identifying a bond distance where the accuracy of CCSD(T) begins to drop for a target molecule. For this, we devise a novel method to detect the accuracy drop of CCSD(T) based on its number of iterations.

Our evaluation using a noise-less quantum computer simulator demonstrates that AAS improves the accuracy to describe the bond breaking processes of four molecules, compared to the cases where CCSD(T) or VQE is solely used. Moreover, AAS reduces the error to calculate bond dissociation energy (BDE) by up to over an order of magnitude.

The organization of the paper is as follows. In \refsec{sec:background}, we first explain QCC, CCSD(T), and VQE. Then, we investigate the accuracy of CCSD(T) and VQE. \refsec{sec:aas} describes the concept and implementation of our proposed AAS technique. \refsec{sec:evaluation} shows the evaluation results of AAS for the bond breaking processes of four molecules. Finally, \refsec{sec:related_work} introduces related work, and \refsec{sec:conclusion} concludes this work.

\section{Background} \label{sec:background}

\subsection{Quantum Chemical Calculations (QCC)} \label{subsec:qcc}

\begin{figure}[t]
    \centering
    \includegraphics[width=\linewidth]{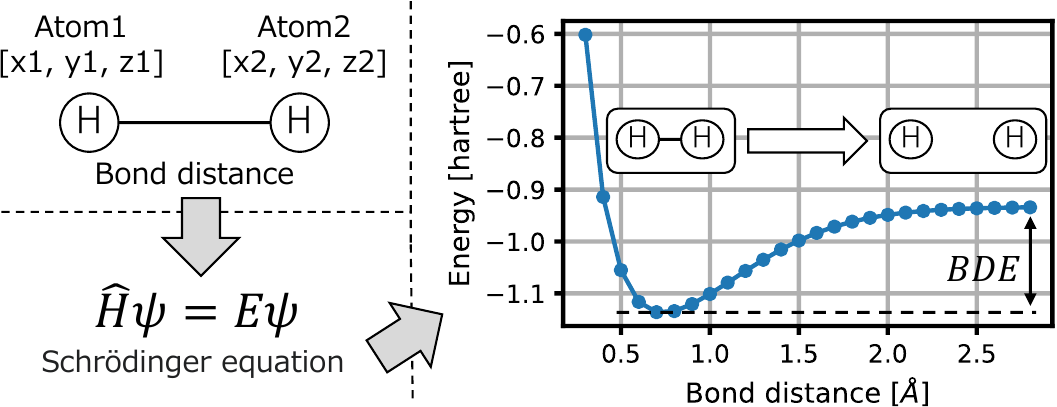}
    \caption{An example to describe the bond breading process of the H2 molecule.}
    \label{fig:pec_example}
\end{figure}

QCC are computational techniques to analyze the characteristics and structures of molecules such as stability and chemical reactions. As shown in the upper left part of \reffig{fig:pec_example}, the input is a molecular structure represented by the coordinates of multiple atoms. The Hamiltonian operator $\hat{H}$ of a target molecule is determined from its structure, and the potential energy $E$ can be calculated by solving the \schr\ equation, $\hat{H}\psi = E\psi$, where $\psi$ is a wave function representing an electron state. However, since exactly solving the equation is not practical, a wide variety of approximate methods with different accuracy levels and computational costs have been developed \cite{Baerends:1973se, Kohn:1965:se, Moller:1934no, David:1999co}. 

The right graph in \reffig{fig:pec_example} shows an example to describe the bond breaking process of the H2 molecule. A potential energy curve (PEC), constructed by calculating potential energies with different bond distances,  represents the change from a stable state at a minimum point to a transition state at a saddle point. The energy difference between saddle and minimum points corresponds to a \emph{bond dissociation energy ($BDE$)}, which is an energy required to break a bond.

\subsection{CCSD(T)} \label{subsec:ccsd_t}

Coupled cluster with singles, doubles, and perturbative triples [CCSD(T)] is regarded as a gold standard algorithm in QCC, because it can achieve a high accuracy with moderate computational cost \cite{Ku:2019ac}. Coupled cluster is a \emph{post-Hartree-Fock} method that represents electron correlation by constructing a multi-electron wave function with cluster operators~\cite{Bartlett:2007co}. \emph{CCSD} explicitly treats single and double excitations with two cluster operators and calculates an electron correlation energy through an iterative process, where the amplitudes are updated repeatedly until the energy difference converges to a threshold~\cite{Solomonik:2014ma}. CCSD(T) further adds the approximate treatment of triple excitations to the energy calculated by CCSD based on the perturbation theory. It is known that CCSD(T) is accurate for stable molecular structures where electron correlation is weak, but not for describing the process of breaking multiple bonds where electron correlation is strong~\cite{Kowalski:2000re}.

\subsection{VQE} \label{subsec:vqe}

\begin{figure}[t]
    \centering
    \includegraphics[width=0.9\linewidth]{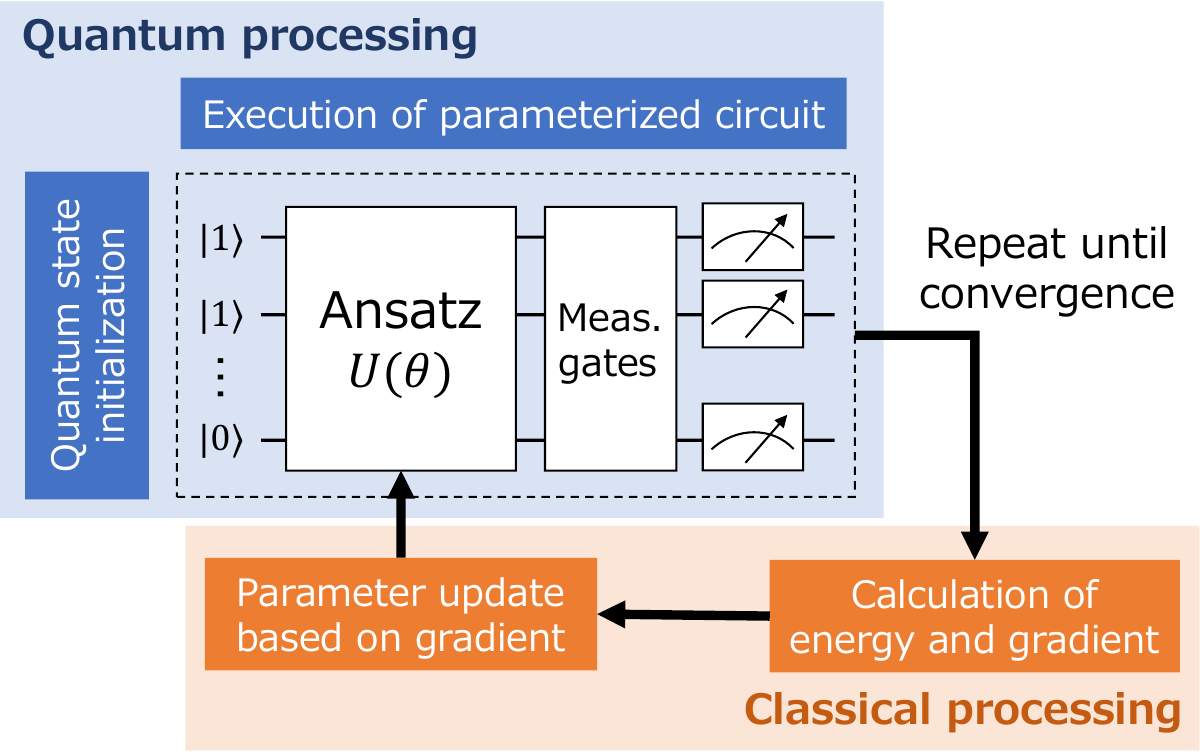}
    \caption{The overview of VQE.}
    \label{fig:vqe_overview}
\end{figure}

With the advent of NISQ computers, variational quantum eigensolver (VQE) is regarded as one of the most promising applications operating on them. It is a variational hybrid quantum-classical eigensolver and can be used to calculate an approximate ground-state energy of a molecule~\cite{Peruzzo:2014va}. \reffig{fig:vqe_overview} illustrates the overview of VQE. It iteratively executes a parameterized quantum circuit on a quantum device and updates its parameters using a classical optimizer until the energy calculated from the measurement result of the quantum circuit converges. The combination of using a shallow parameterized quantum circuit and using a classical optimizer enables VQE to be executed on NISQ computers where coherent time is short. With a well-known mapping method called \emph{Jordan-Wigner}~\cite{Fradkin:1989jo}, the number of qubits of a quantum circuit used in VQE corresponds to the number of spin orbitals of a target molecule, and the measurement value of each qubit ($|0\rangle$ or $|1\rangle$) represents whether an electron occupies the corresponding spin orbital or not. 

A parameterized quantum circuit used in VQE is constructed based on a trial wave function called \emph{ansatz}. The accuracy and computational cost of VQE heavily depend on an ansatz; thus, various types of ansatze have been proposed~\cite{Ryabinkin:2018qu,Grimsley:2019ad,Lee:2019ge,Fan:2021ci,Lan:2022am}. In particular, \emph{unitary coupled-cluster singles and doubles (UCCSD)} is a chemistry-inspired ansatz which can achieve a high accuracy in QCC~\cite{Ku:2019ac,Gonthier:2022me}. In addition, the type of a classical optimizer also significantly affects the accuracy and the number of iterations required until convergence; thus, various types of classical optimizers have been evaluated~\cite{Bonet:2021pe,Skogh:2023ac}. \emph{Sequential least squares programming (SLSQP)} is a representative optimizer in the absence of noise, while \emph{simultaneous perturbation stochastic approximation (SPSA)} is a well-known optimizer robust to noise~\cite{Skogh:2023ac}.

\subsection{Accuracy of CCSD(T) and VQE} \label{subsec:motivation}

\begin{figure}[t]
    \centering
    \includegraphics[width=\linewidth]{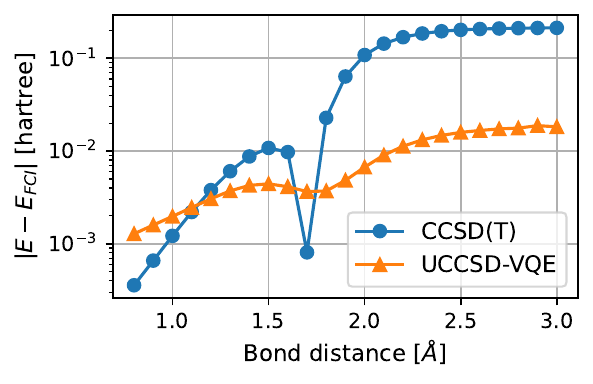}
    \caption{The energy error from FCI for N2/STO-3G with different bond distances.}
    \label{fig:error_N2}
\end{figure}

We investigate the accuracy of CCSD(T) implemented in PySCF~\cite{pyscf} and UCCSD-VQE executed using a quantum computer simulator, \emph{AerSimulator}, in Qiskit~\cite{Qiskit} without a noise model (see \refsec{subsec:setup} for the detailed experimental setup). \reffig{fig:error_N2} shows the energy error of CCSD(T) and UCCSD-VQE from the exact energy calculated with the \emph{full configuration interaction (FCI)} method for the N2 molecule within the STO-3G basis set (denoted as N2/STO-3G) according to different bond distances. We can see that the accuracy advantage between CCSD(T) and UCCSD-VQE differs depending on bond distances. CCSD(T) achieves a lower error than UCCSD-VQE at 0.8 to 1.1~\ang, whereas UCCSD-VQE is more accurate at longer bond distances except 1.7~\ang. For the N2 molecule having a triple bond, it is known that CCSD(T) is not accurate at long bond distances because it cannot accurately represent strong electron correlation~\cite{Kowalski:2000re}. In contrast, UCCSD-VQE represents it more accurately than CCSD(T).

\section{Automatic Algorithm Switching} \label{sec:aas}

In this work, we propose an \emph{automatic algorithm switching (AAS)} technique for accurate ground-state energy calculation with the different bond distances of a target molecule. As shown in \reffig{fig:aas_overview}, AAS automatically switches CCSD(T) and VQE according to bond distances by identifying a bond distance where the accuracy of CCSD(T) begins to drop for a target
molecule.

\subsection{Switch Point Identification}

The key of our proposed technique is to appropriately identify a switch point between CCSD(T) and VQE. It is desirable to switch from CCSD(T) to VQE at a bond distance where the accuracy of CCSD(T) begins to drop, although this is not a trivial task because such a bond distance is different depending on the characteristics of molecules. The accuracy of CCSD(T) can be known exactly by comparing the energy calculated with CCSD(T) and the exact energy calculated with FCI. However, FCI is not available for practical molecules due to its exponential computational and memory requirements~\cite{Sugisaki:2021qu}. 

\begin{figure}[t]
    \centering
    \subfloat[LiH/STO-3G]{\includegraphics[width=0.9\linewidth]{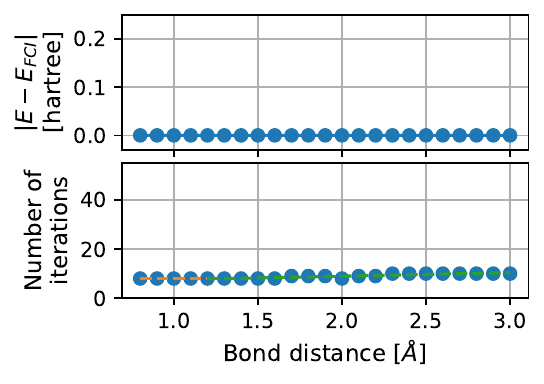}
    \label{subfig:error_niters_LiH}}
    \\
    \subfloat[N2/STO-3G]{\includegraphics[width=0.9\linewidth]{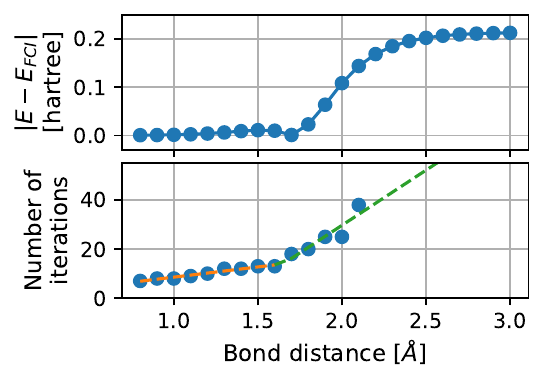}
    \label{subfig:error_niters_N2}}
    \caption{The energy error (upper) and number of iterations (lower) of CCSD(T) with different bond distances. The orange and green dotted lines are fitted to the number of iterations with piecewise linear regression, where the points over 50 iterations are excluded.}
    \label{fig:error_niters}
\end{figure}

Since CCSD(T) performs an iterative process to calculate an electron correlation energy, we focus on the relationship between the accuracy and number of iterations of CCSD(T). \reffig{fig:error_niters} shows the energy error from FCI and number of iterations of CCSD(T) for LiH and N2 molecules within the STO-3G basis set with different bond distances. For LiH/STO-3G in \reffig{subfig:error_niters_LiH}, the energy error is always very low regardless of bond distances, and the number of iterations is almost constant. On the other hand, for N2/STO-3G in \reffig{subfig:error_niters_N2}, the energy error increases sharply from 1.8~\ang, and the number of iterations also does so from 1.7~\ang. The iterative process of CCSD(T) converges with around ten iterations when the energy error is low, whereas the number of iterations increases linearly as the energy error increases. These results demonstrate that the sharp increase in the number of iterations appropriately represents the accuracy drop of CCSD(T).

To identify a switch point from CCSD(T) to VQE for a target molecule, we fit the number of iterations of CCSD(T) according to bond distances with piecewise linear regression (PWLR)~\cite{pwlf}. In the lower graphs of \reffig{fig:error_niters}, the orange and green dotted lines represent the two fitted lines. For LiH/STO-3G in \reffig{subfig:error_niters_LiH}, the slopes of both the lines are almost zero. On the other hand, for N2/STO-3G in \reffig{subfig:error_niters_N2}, their slopes are both positive and the right green line has a higher slope than the left orange line. We regard a break point between the two lines as a switch point from CCSD(T) to VQE, because the break point corresponds to a bond distance where the accuracy of CCSD(T) may begin to drop.

\subsection{Implementation}

\begin{algorithm}[t]
    \caption{Switch point identification}
    \label{algo:spi}
    \begin{algorithmic}[1]
        \REQUIRE mol, distances, niters\_thr
        \STATE niters $\leftarrow$ runall\_ccsd\_t(mol, distances)
        \STATE break\_point, slopes \\ $\leftarrow$ pwlr\_fit(distances, niters, niters\_thr)
        \IF {0 $<$ slopes[0] $<$ slopes[1]}
            \STATE switch\_point $\leftarrow$ break\_point
        \ELSE
            \STATE switch\_point $\leftarrow$ -1
        \ENDIF
        \RETURN switch\_point
    \end{algorithmic}
\end{algorithm}

On the basis of the above finding, we implement a function to identify a switch point between CCSD(T) and VQE as shown in Algorithm~\ref{algo:spi}. The inputs are the information of a target molecule such as its geometry and basis set ({\tt mol}), the range of bond distances ({\tt distances}), and a threshold of the number of iterations ({\tt niters\_thr}). At first, CCSD(T) is executed for the target molecule at all the bond distances included in {\tt distances}, and the numbers of iterations are recorded (line~1). Then, the break point and slopes of two fitted lines are obtained by applying PWLR to the recorded numbers of iterations according to the bond distances (line~2). PWLR only fits the numbers of iterations less than {\tt niters\_thr}. In this work, {\tt niters\_thr} is set to 50, which is the default maximum iterations of CCSD(T) in PySCF. If the slopes of the two fitted lines are both positive and the second slope is higher than the first one, the switch point is set to the break point (lines 3-4). Otherwise, the switch point is set to ``-1'' for disabling the automatic algorithm switching (lines 5-6). Finally, the switch point is returned (line 8).

Our proposed AAS technique switches CCSD(T) and VQE based on the switch point obtained from Algorithm~\ref{algo:spi} and a target bond distance. CCSD(T) is selected if a target bond distance is shorter than the switch point; otherwise VQE is selected. If the switch point is ``-1'', CCSD(T) is always selected (i.e., VQE is never selected). 

\section{Evaluation} \label{sec:evaluation}

In this section, we compare the accuracy to describe the bond breaking processes of four molecules between CCSD(T), VQE, and our proposed AAS technique. We first explain the experimental setup and then show evaluation results.

\subsection{Experimental Setup} \label{subsec:setup}

In this work, we target four molecules within the STO-3G basis set: LiH, N2, C2, and CO. The three molecules except LiH have double or triple bonds, leading to the low accuracy of CCSD(T) at long bond distances. The bond breaking process of each molecule is described by constructing a potential energy curve (PEC) across bond distances from 0.8 to 3.0 \ang\ at the step of 0.1 \ang. The accuracy of CCSD(T), VQE, and AAS is evaluated by comparing an energy calculated with each method to an exact energy calculated with FCI. For all experiments in this work, we use a server containing two Xeon\textregistered\ Gold 6240M processors and 384~GB of DRAM.

\begin{table}[t]
    \caption{The parameter settings of VQE.}
    \label{tab:vqe_setting}
    \centering
    \begin{tabular}{rl}
        \hline
        {\bf Parameter} & {\bf Setting} \\
        \hline
        \hline
        Ansatz & UCCSD \\
        Classical optimizer & SLSQP \\
        Maximum iterations & 100 \\        
        Core freezing & Enabled \\
        Qubit converter & Jordan Wigner \\
        Z2 symmetry reduction & Enabled \\
        Noise & None\\
        \hline
    \end{tabular}
\end{table}

PySCF~\cite{pyscf} is used to run CCSD(T) and FCI, while Qiskit~\cite{Qiskit} is used to run VQE. PySCF is a Python-based open-source QCC framework that supports various classical algorithms. We set the maximum number of iterations of CCSD(T) to 10,000. Qiskit is an open-source quantum computing framework developed by IBM Research. In this work, we run VQE using a quantum computer simulator, \emph{AerSimulator}, in Qiskit without a noise model (i.e., noise-less simulation). \reftab{tab:vqe_setting} summarizes the parameter settings of VQE. We select the UCCSD ansatz and SLSQP optimizer to achieve high accuracy in the noise-less simulation. Core freezing is applied for each molecule to reduce the number of qubits without degrading the VQE accuracy significantly. The Hamiltonian of each molecule is converted to qubits with the Jordan Wigner method~\cite{Fradkin:1989jo} and tapered with the Z2 symmetry reduction method~\cite{bravyi:2017ta}. Consequently, the numbers of qubits are 6, 12, 12, and 13 for LiH, N2, C2, and CO, respectively. Moreover, we use the \emph{pwlf} Python library~\cite{pwlf} to apply piecewise linear regression in Algorithm~\ref{algo:spi}.

\subsection{Evaluation Results}

\begin{figure*}[t]
    \centering
    \includegraphics[width=\linewidth]{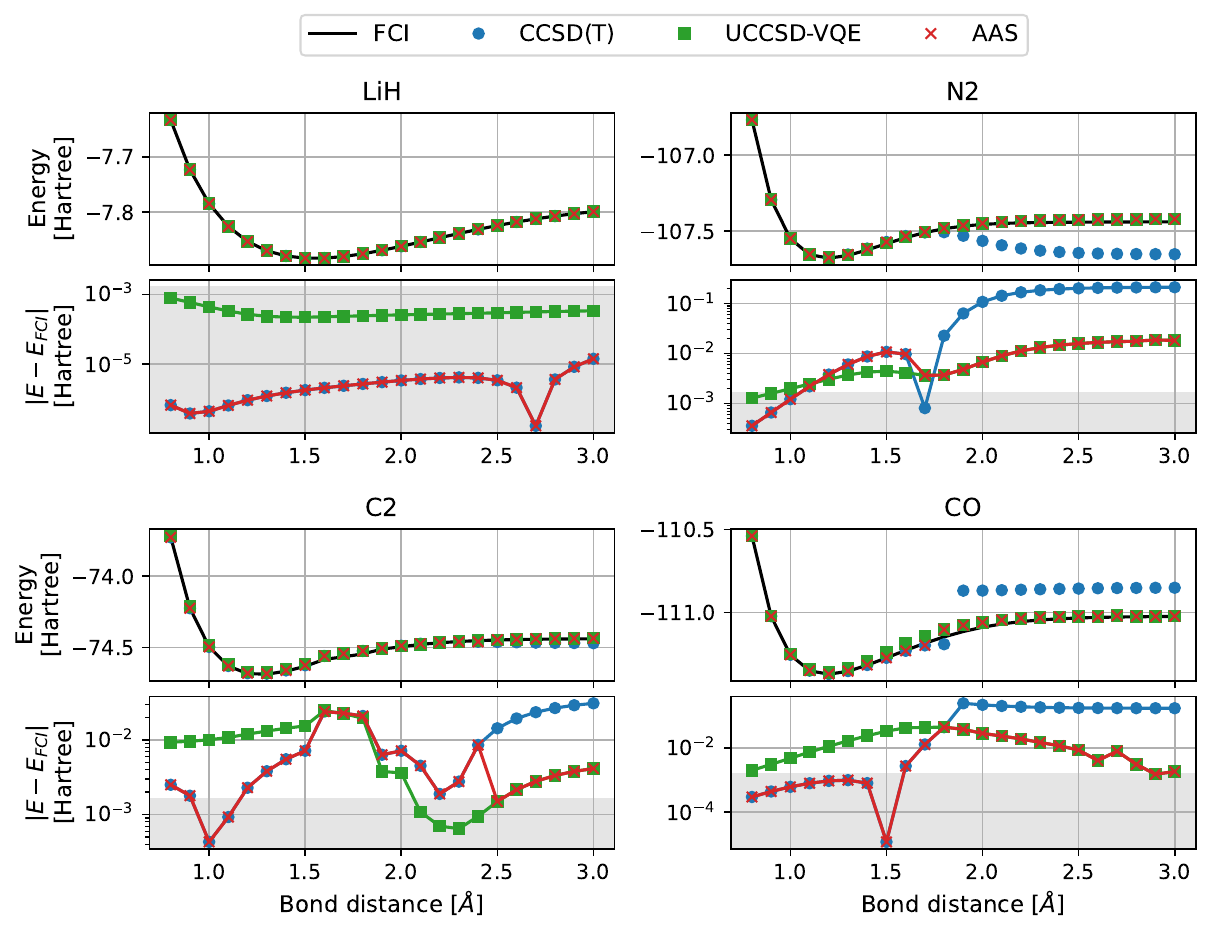}
    \caption{The potential energy curve (upper) and energy error from FCI (lower) for four molecules within the STO-3G basis set. The gray shade in each lower graph represents the chemical accuracy.}
    \label{fig:pec_error_all}
\end{figure*}

First, we evaluate the accuracy to describe bond breaking processes by comparing PECs constructed with CCSD(T), UCCSD-VQE, and AAS. \reffig{fig:pec_error_all} plots the PECs and energy errors from FCI for the four molecules. The black line in each upper graph shows the exact PEC constructed with FCI as an accuracy criterion. The gray shade in each lower graph represents the \emph{chemical accuracy} (i.e., energy error within 1.6E-3 Hartree), which is an important accuracy criterion to reproduce chemical experimental results.

For LiH having a single bond, all the three methods successfully construct accurate PECs and always achieve the chemical accuracy. In this case, CCSD(T) always achieves much lower errors than UCCSD-VQE. Since the number of iterations of CCSD(T) is almost constant as shown in \reffig{subfig:error_niters_LiH}, AAS selects only CCSD(T), and thus the energy errors of CCSD(T) and AAS (the blue and red lines) completely overlap in the lower graph. In contrast, for the other three molecules having double or triple bonds, CCSD(T) achieves lower errors than UCCSD-VQE at short bond distances, but it becomes less accurate than UCCSD-VQE and fails to construct accurate PECs at long bond distances. This is because CCSD(T) cannot represent strong electron correlation accurately~\cite{Kowalski:2000re}. On the other hand, UCCSD-VQE successfully constructs accurate PECs even at long bond distances because of the UCCSD ansatz and its variational property~\cite{Mizukami:2020or}. For the three molecules, AAS takes the accuracy advantages of both CCSD(T) and UCCSD-VQE by identifying an appropriate switching point and switching them according to bond distances.

Only in terms of the accuracy to describe bond breaking processes, we may be able to achieve the comparable accuracy to AAS just by switching from CCSD(T) to UCCSD-VQE at a fixed point (e.g., at 1.8~\ang\ for the four molecules targeted in this work). However, this simple approach wastes the scarce resources of current quantum computers. For instance, it needs 52 runs of UCCSD-VQE in total to construct the PECs of the four molecules (13 runs across 1.8~\ang\ to 3.0~\ang\ for each molecule). In contrast, AAS reduces the total runs of UCCSD-VQE to 33 ($= 0+14+6+13$ for LiH, N2, C2, and CO). The saving of scarce quantum computing resources by the hybrid use of classical algorithms like CCSD(T) and emerging quantum algorithms like VQE is another important aspect of AAS.

\begin{figure}[t]
    \centering
    \includegraphics[width=\linewidth]{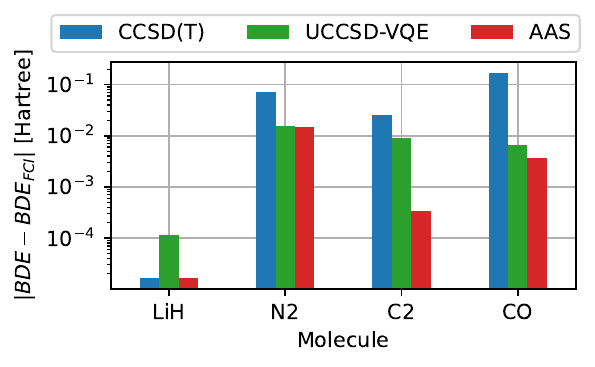}
    \caption{The error of bond dissociation energy ($BDE$) from FCI for four molecules within the STO-3G basis set.}
    \label{fig:bde_error}
\end{figure}

Next, we evaluate the accuracy to calculate the bond dissociation energy ($BDE$) of each molecule. As shown in the right graph of \reffig{fig:pec_example}, BDE is calculated as the energy difference between the saddle and minimum points on a PEC. \reffig{fig:bde_error} plots the error of BDE calculated with CCSD(T), UCCSD-VQE, and AAS from the exact BDE calculated with FCI. For LiH, CCSD(T) achieves a much lower error than UCCSD-VQE because it constructs a more accurate PEC. Since AAS selects only CCSD(T) in this case, it achieves the same accuracy to CCSD(T). For N2, UCCSD-VQE  outperforms CCSD(T) because CCSD(T) is not accurate at long bond distances. AAS is comparable to UCCSD-VQE in this case, because the errors of CCSD(T) and UCCSD-VQE at the minimum point is comparable. More interestingly, AAS achieves an order of magnitude lower errors than CCSD(T) and/or UCCSD-VQE for C2 and CO. This is because the accuracy advantage between CCSD(T) and UCCSD-VQE differs significantly depending on bond distances, and AAS takes the advantages of both of them by appropriately switching them.

\section{Related Work}
\label{sec:related_work}

To our knowledge, this work is the first to present an algorithm switching technique for the accurate calculation of molecular ground-state energy with different bond distances. In this section, we introduce prior work regarding accurate ground-state energy calculation.

Classical algorithms have been improved in terms of accuracy and/or computational costs. Kowalski and Piecuch show the low accuracy of CCSD(T) for the bond breaking process of the N2 molecule due to strong electron correlation~\cite{Kowalski:2000re}. CCSD(T) has an $O(N^7)$ computational cost, where $N$ is a problem size; thus, it cannot be used for large-scale molecules. \emph{Domain-based local pair natural orbital} CCSD(T) [DLPNO-CCSD(T)] is a promising algorithm to reduce the computational cost of CCSD(T) while sustaining its potential accuracy~\cite{Riplinger:2013ef,Minenkov:2015ac,Sandler:2021ac}. \emph{Selected CI (SCI)} is a well-known approach to reduce the computational costs of configuration interaction methods including FCI by selecting only critical electron configurations. Coe applies an artificial neural network model to SCI~\cite{Coe2019:ma}, while Li \etal combine SCI with the perturbation theory~\cite{Li:2020ac}. Aroeira \etal augment CCSD(T) with an adaptive SCI method to represent strong electron correlation~\cite{Aroeira:2021co}. These DLPNO-CCSD(T) and SCI methods can be used as candidates selected in AAS.

Since VQE is a promising algorithm for accurate ground-state energy calculation, a wide variety of approaches have been proposed to improve its accuracy and/or computational cost. Although the UCCSD ansatz can achieve a high accuracy in the noise-less simulation as we have shown in this paper, UCCSD-based quantum circuits have the impractical depth for current NISQ computers and suffer from significant noise errors. Therefore, the various types of more light-weight ansatze have been proposed: {\it QCC} \cite{Ryabinkin:2018qu}, {\it k-UpCCGSD} \cite{Lee:2019ge}, {\it CHC} \cite{Ollitrault:2020ha}, and {\it ADAPT} \cite{Grimsley:2019ad,Fan:2021ci,Lan:2022am}. In addition, quantum circuit pruning techniques have been proposed to reduce the number of gates and depth of quantum circuits used in VQE~\cite{Sim:2021ad, Wang:2022qu, Imamura:2023of}. While we evaluate UCCSD-VQE using a noise-less quantum computer simulator in this work, we need to thoughtfully consider the applications of the above approaches to achieve accurate VQE calculation on real NISQ computers. 

The \emph{density matrix embedding theory (DMET)} can be used to divide molecules into multiple smaller fragments and combined with VQE to calculate the energy of each fragment~\cite{Kawashima:2021op}. Shang \etal combine DMET with VQE executed using a tensor network-based quantum computer simulator to calculate the energies of large-scale molecules on a supercomputer~\cite{Shang:2022la}. The number of gates and depth of quantum circuits used in VQE become higher for larger molecules, leading to larger noise errors to VQE calculation on real NISQ computers. DMET should be effective to mitigate noise errors because it can reduce a problem size for VQE.

\section{Conclusion} \label{sec:conclusion}

In this work, we propose an automatic algorithm switching (AAS) technique to accurately calculate the ground-state energies of a target molecule with different bond distances. We focus on the relationship between the accuracy and number of iterations of CCSD(T) and present a novel approach that identifies an appropriate switching point between CCSD(T) and VQE based on the number of iterations of CCSD(T). Our evaluation in terms of the accuracy to describe the bond breaking processes of four molecules demonstrates that compared to CCSD(T) and UCCSD-VQE, our proposed AAS technique successfully constructs more accurate potential energy curves and reduces the error of bond dissociation energy by up to over an order of magnitude.

As our future work, we will evaluate the accuracy of VQE on real NISQ computers and try to achieve accurate ground-state energy calculation by applying AAS to switch CCSD(T) and such noisy VQE. This is a challenging task due to the significant noise errors on VQE calculation. To minimize the noise errors, we need to thoughtfully construct an accurate and light-weight ansatz, select a noise-robust classical optimizer, and apply various noise mitigation methods. 

\bibliographystyle{IEEEtran}
\bibliography{references}

\end{document}